\documentclass{cernrep} 
\usepackage{texnames}
\usepackage[T1]{fontenc}
\pdfoutput=1

\newcommand{\babar}{\mbox{\ensuremath{{\displaystyle B}\!{\scriptstyle A}
{\displaystyle B}\!{\scriptstyle AR}}}}

\begin{document}
\title{SVD-based unfolding: implementation and experience}
 
\author{Kerstin Tackmann, Andreas H\"ocker}

\institute{CERN, Geneva, Switzerland}

\maketitle 

\begin{abstract}
With the first year of data taking at the LHC by the experiments, unfolding
methods for measured spectra are reconsidered with much interest. Here, we
present a novel ROOT-based implementation of the Singular Value Decomposition
approach to data unfolding, and discuss concrete analysis experience
with this algorithm.
\end{abstract}
 
\section{Introduction}
 
The measured spectrum of a physical observable is usually distorted by 
detector effects, such as finite resolution and limited acceptance. A 
comparison of the measured spectrum with theoretical predictions requires a 
removal of these effects to obtain the true, underlying physical spectrum, or 
the folding of these effects into the theoretical prediction. Unfolding 
methods provide ways for correcting the measured distributions, where the 
difficulty lies in the statistical instability of the inversion problem, 
requiring regularization. One widely used method is based on a singular value 
decomposition (SVD) of the detector response matrix~\cite{bib:svd}.

The unfolding problem can be formulated as a matrix equation, 
$\hat{A}_{ij}x_j = b_i$, where $x$ is the true, physical distribution, $b$ 
the measured distribution. $\hat{A}_{ij}$ is the probability for an event 
generated in bin $j$ to be reconstructed in bin $i$ and as such, $\hat{A}$ 
describes finite resolution and inefficiencies and can be obtained from the 
simulation (or appropriate control samples). The singular value decomposition 
of $\hat{A}$ serves both for shedding light on the underlying instability of 
the problem, as well as for providing a solution. Small singular values, which 
are often present in detector response matrices, are found to greatly enhance 
statistical fluctuations in the measured distribution. A suitably chosen 
regularization procedure dampens the enhanced fluctuations. Rewriting the 
above equation to $A_{ij}w_j = b_i$, where $A_{ij}$ now contains numbers of 
events rather than probabilities, and $w$ describes the ratio between the 
desired physical distribution and the underlying true distribution in the 
simulation (for example), allows for a better treatment of the statistical 
uncertainties in the detector matrix. At the same time, this allows for a 
physically motivated regularization via a discrete minimum-curvature condition 
on the ratio of the unfolded distribution and a simulated truth distribution, 
which corresponds to retaining the statistically significant contributions of 
$w$, shown to be related to the larger singular values in the decomposition of
$A$.

This note presents a \texttt{C++} implementation of the SVD-based unfolding, 
discusses analysis experience with this algorithm, and provides a comparison 
to the iterative dynamically stabilized unfolding method (IDS)~\cite{bib:ids}
for a concrete example.

\section{ROOT-based implementation}

A \texttt{C++} implementation of the SVD-based unfolding is provided by 
\texttt{TSVDUnfold}, which is part of the ROOT analysis 
framework~\cite{bib:root} as of version 5.28. It can also be used through the 
\texttt{RooUnfold} framework~\cite{bib:roounfold}, which is based on ROOT and 
comes with additional functionality.

\texttt{TSVDUnfold} provides access to the singular values of the detector 
response matrix and to the distribution of the $|d_i|$ (see 
Ref.~\cite{bib:svd}), which help to properly set the regularization strength 
parameter in the unfolding. \texttt{TSVDUnfold} also allows to propagate 
covariance 
matrices of the measured spectrum through the unfolding using pseudo 
experiments. In addition it provides the covariance matrix of the unfolded 
spectrum related to finite statistics in the simulation sample (or control 
sample) that is used to determine the detector response matrix, also making 
use of pseudo experiments.

More recently, \texttt{TSVDUnfold} has been extended to also provide the 
regularized covariance matrix and the inverse covariance matrix (not 
regularized) computed during the unfolding (see Eqs.~(52,53) in
Ref.~\cite{bib:svd}). In addition, the new version of \texttt{TSVDUnfold} 
implements the internal rescaling of the unfolding equations making use of the 
full covariance matrix of the measured spectrum (see Eq.~(34) in 
Ref.~\cite{bib:svd}) rather than only its diagonal elements.

\section{Covariance matrices}
\label{sec:covmats}

\begin{figure}
\centering{\includegraphics[width=.42\linewidth]{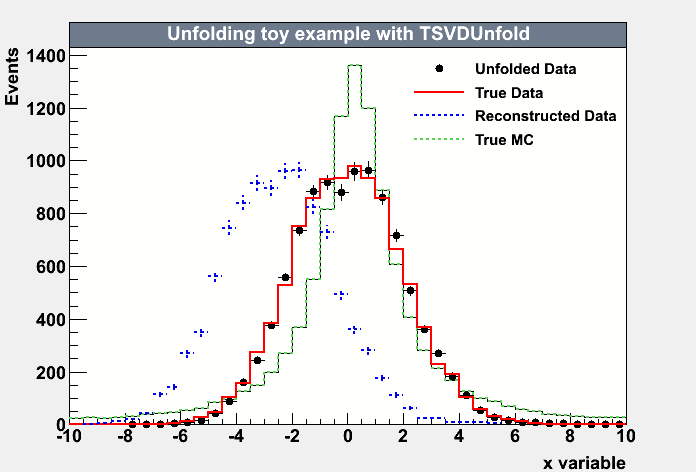}
\includegraphics[width=.42\linewidth]{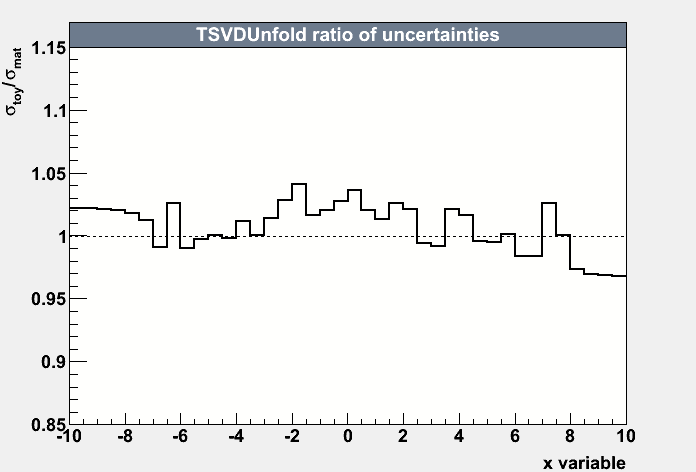}\\
\includegraphics[width=.42\linewidth]{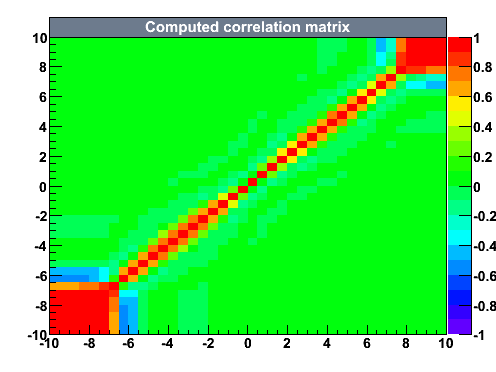}
\includegraphics[width=.42\linewidth]{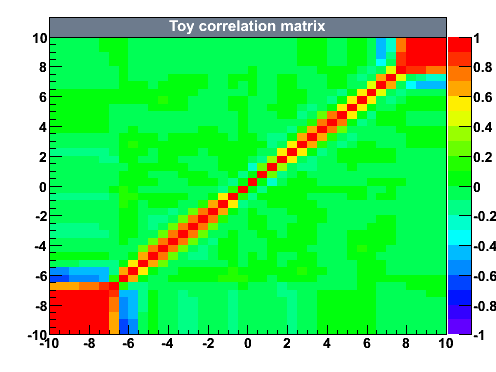}}
\caption{Unfolding toy example. (Left top) Reconstructed and unfolded toy 
data, as well as the truth distributions for toy data and toy simulation. 
(Right top) Ratio of diagonal errors obtained by pseudo experiments and 
computed during the unfolding. (Left bottom) Correlation matrix on the 
unfolded spectrum as computed during the unfolding and (Right bottom) as 
obtained from pseudo experiments.}
\label{fig:covmats}
\end{figure}

The covariance matrices of the unfolded spectrum as computed during the 
unfolding and as obtained from pseudo experiments, respectively, have been 
compared for a toy example (see Fig.~\ref{fig:covmats}) and have been found
in good agreement. The uncertainties (taken from the diagonal elements of the 
covariance matrices) provided by the two methods agree to better than $4\%$ 
and the correlations are well-reproduced. Even in the case of non-optimal 
regularization, the two methods provide compatible results: the uncertainties
obtained with the two methods have been found to agree within $6\%$ ($11\%$) 
for a strongly under- (over-) regularized unfolding, with compatible 
correlation patterns.

\section{Experience with SVD-unfolding in \babar}

The SVD-based unfolding has been used in numerous data analyses over the past 
$15$ years, among which is the unfolding of the hadronic mass spectrum in 
inclusive, charmless, semileptonic $B$-meson decays, $B \to X_u\ell\nu$, at 
the \babar\ experiment~\cite{bib:babar}. Due to the nature of the measured 
spectrum, its unfolding and in particular the determination of the appropriate 
regularization required careful studies. The relatively low statistics of 
estimated $1027$ signal events and the subtraction of the dominant 
$B\to X_c\ell\nu$ backgrounds result in sizable statistical and systematic 
uncertainties. The size of the bins has been chosen to equal the hadronic 
mass resolution in signal events. Due to the large uncertainty in the 
reconstruction efficiency of the tagging method used, which results in a 
significantly better hadronic mass resolution, the unfolded spectrum is 
normalized to unit area, which results in increased bin-by-bin correlations.

The regularization has been determined with the use of pseudo experiments, 
where the toy data and toy simulation distributions and detector response 
differ in the assumed value of the $b$-quark mass, which determines the shape 
of the inclusive hadronic mass spectrum and is one of the primary results of
the analysis. The regularization has been chosen such that the unfolding bias 
in the spectral moments of the unfolded spectrum, which are directly related 
to the $b$-quark mass, is small compared to their statistical errors.

A few observations have been made which are of relevance for the unfolding of 
spectra with sizable uncertainties. The unfolding gains stability when the 
internal rescaling of the equations that is performed by the SVD-based 
unfolding (see Ref.~\cite{bib:svd}) takes both the statistical and the 
systematic uncertainties into account, since this provides a better estimate 
of how well the different regions of the measured distribution are known. 
Moreover, the propagation of the covariance matrices of the measured spectrum
to the unfolded spectrum shows a more linear behavior and is less likely to be 
affected by instabilities in the unfolding in the presence of sizable 
uncertainties when the covariance matrices related to different sources of 
uncertainties are propagated separately and then combined.

\begin{figure}
\centering{\includegraphics[width=.44\linewidth]{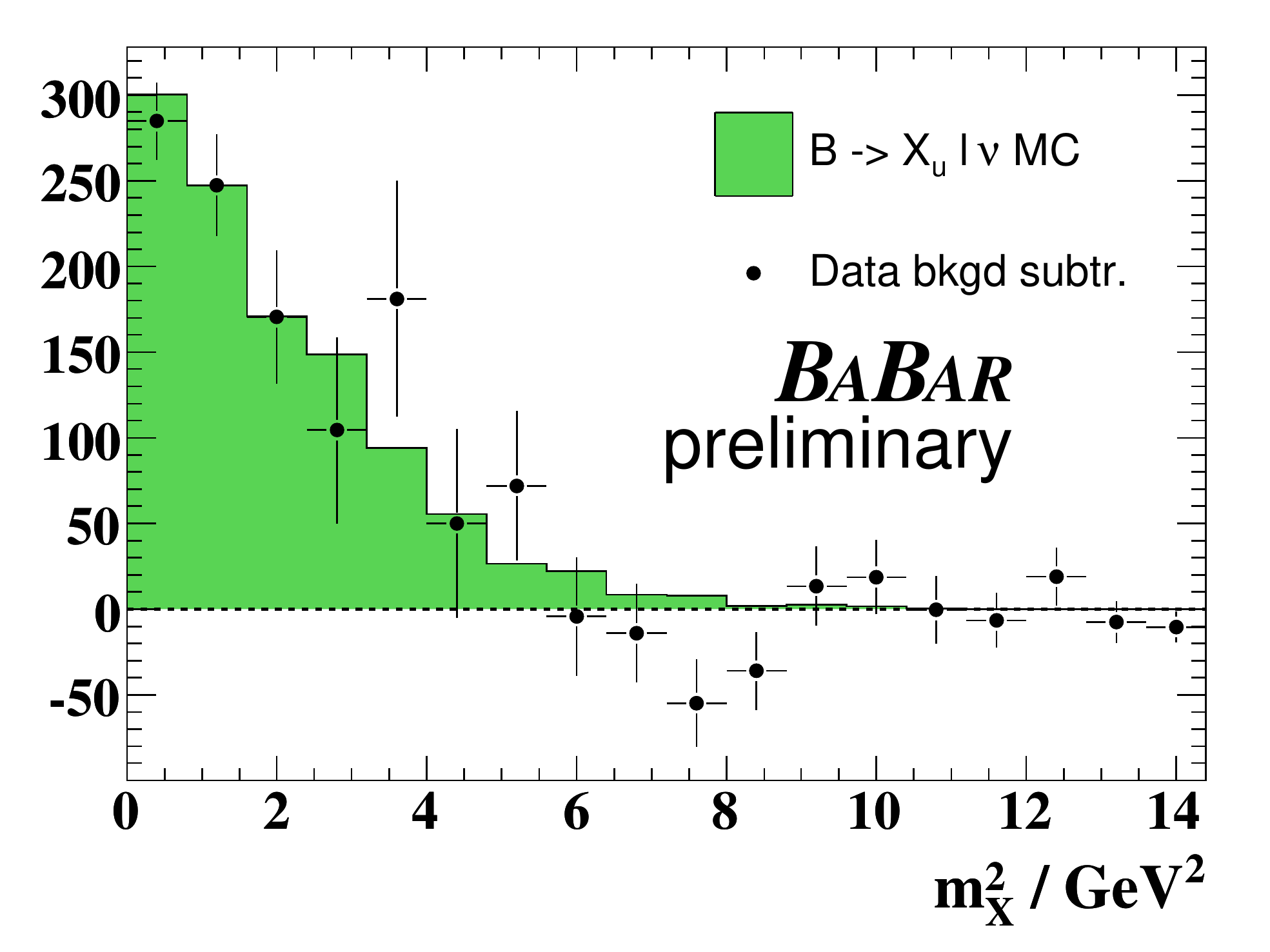}
\includegraphics[width=.465\linewidth]{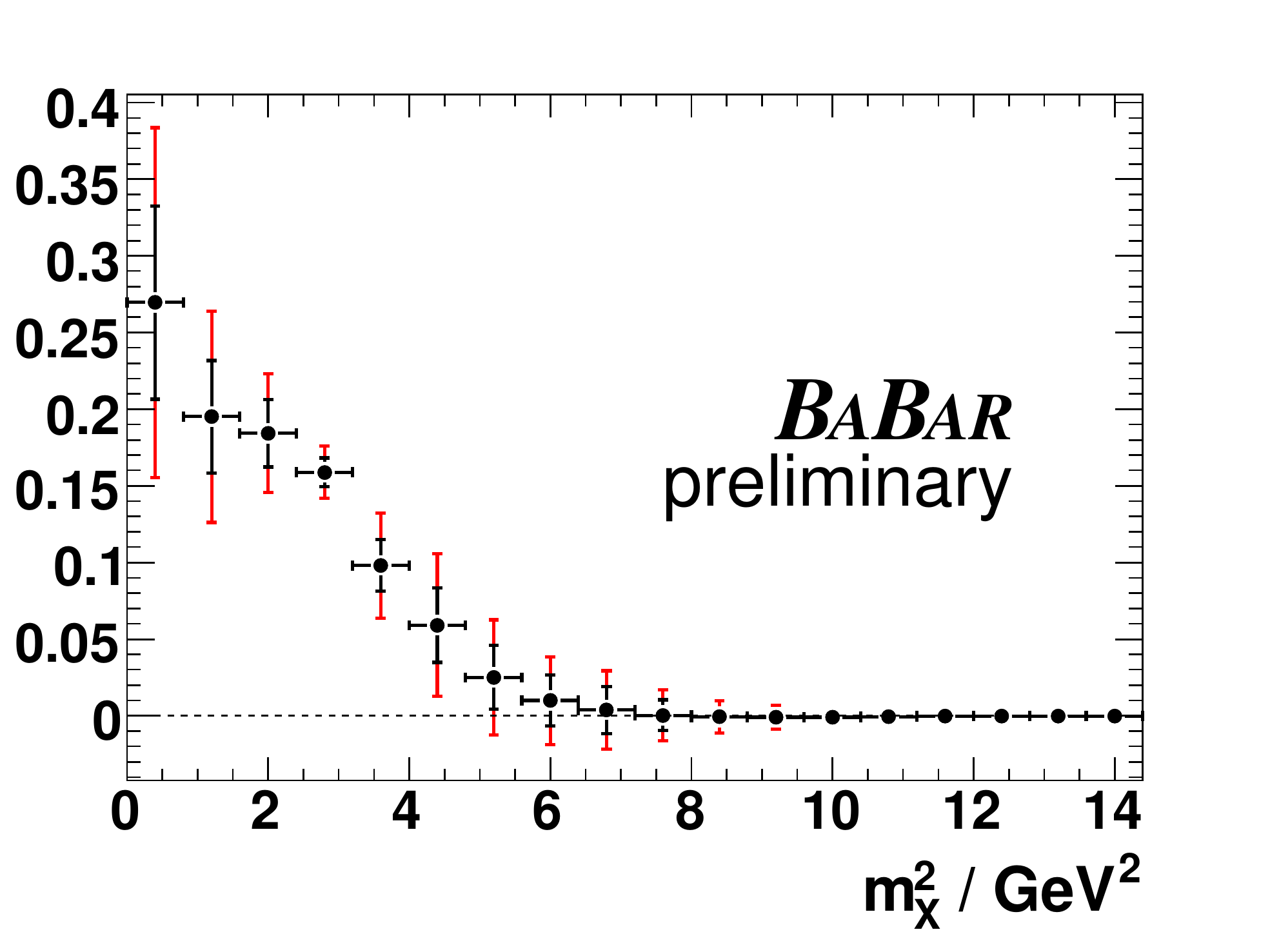}}
\caption{Unfolding example from the \babar\ experiment. (Left) Measured
hadronic mass spectrum (statistical uncertainties only) and signal Monte Carlo 
simulation in $B\to X_u\ell\nu$ decays. (Right) Normalized unfolded
hadronic mass spectrum with total (outer error bars) and statistical 
(inner error bars) uncertainties.}
\label{fig:babar}
\end{figure}

\section{Comparison to interative dynamically stabilized unfolding}

It is instructive to compare the results of different unfolding methods for 
the same example. Here, we present the results for the toy example of 
Sect.~\ref{sec:covmats}, using both SVD-based and IDS [2] unfolding.

\begin{figure}
\centering{\includegraphics[width=.45\linewidth]{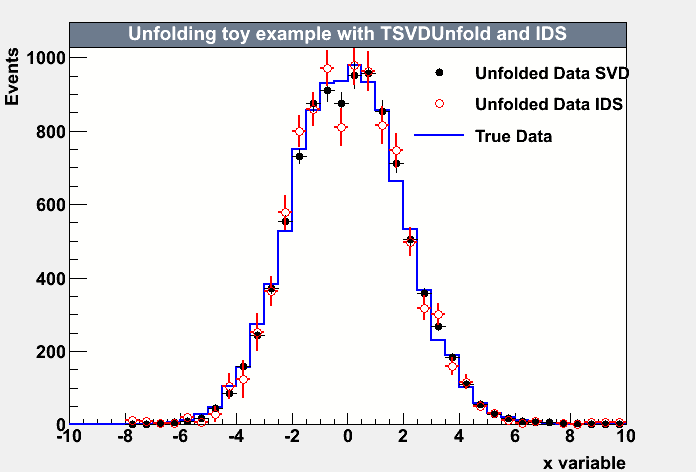}
\includegraphics[width=.45\linewidth]{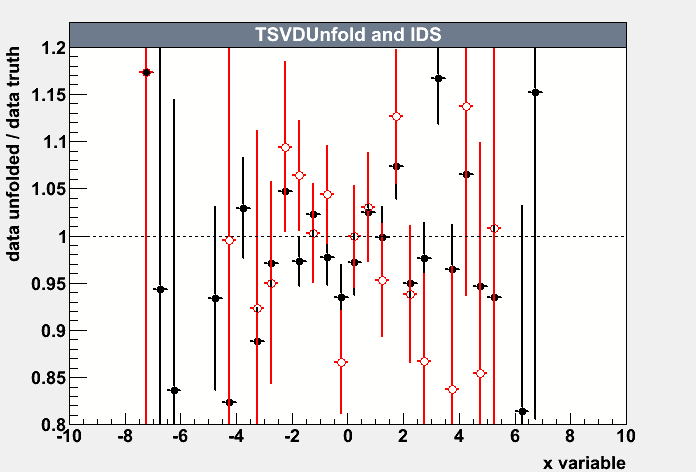}}
\caption{Unfolding toy example. (Left) The results obtained with SVD-based 
unfolding and IDS unfolding are compared to the true toy distribution. 
(Right) The ratio of the unfolded distributions and the truth distribution.}
\label{fig:svdandids}
\end{figure}

The regularization for the two methods has been determined independently. For 
the SVD-based unfolding, the distribution of the $|d_i|$ has been used to 
chose the regularization ($k = 16$). For the IDS unfolding, it has been 
determined using the toy data as well as the reconstructed improved toy
simulation distributions. The unfolding results can be seen in 
Fig.~\ref{fig:svdandids}. Neither unfolding result shows any obvious bias with 
the chosen regularization. However, the result of the IDS unfolding shows 
somewhat larger fluctuations around the true distributions as well as larger 
uncertainties, which points to a looser regularization than that used for the 
SVD-based unfolding. In addition, the observed pattern in the bin-by-bin 
correlations is very different. The result of the SVD-based unfolding shows 
positive correlations between neighbouring bins, negative correlations in the 
medium range, and very small correlations in the long range. The result of the 
IDS unfolding in general shows smaller correlations, and neighbouring bins 
tend to be anti-correlated. In general, the stronger the regularization, the 
larger and broader are the positive correlations between adjacent bins. The 
difference in the correlations observed between the SVD-based and IDS 
unfolding results are due to the stronger regularization in the SVD-based 
unfolding, which is also apparent in the smaller diagonal errors.

\section{Summary}

\texttt{TSVDUnfold} provides a \texttt{C++} implementation of the SVD-based 
unfolding algorithm and is available as part of the ROOT analysis framework. 
Recently, it has been improved to take into account bin-by-bin correlations 
in the measured spectrum. SVD-based unfolding has been successfully used in 
many data analyses and a concrete example from the \babar\ experiment has been 
presented, along with observations that are of relevance for unfolding spectra 
which are subject to large uncertainties. In addition, unfolding results for 
a toy example have been compared using SVD-based and IDS unfolding.

\section*{Acknowledgements}
The authors wish to thank Vakhtang Kartvelishvili for discussions and advice 
related to the SVD-based unfolding, as well as Heiko Lacker for the 
collaboration in the implementation of \texttt{TSVDUnfold}. We furthermore 
wish to thank Bogdan Malaescu for general discussions on unfolding and 
providing the unfolded example spectrum using IDS unfolding. Thanks to 
Lorenzo Moneta \texttt{TSVDUnfold} is now distributed as part of the ROOT 
analysis framework, and thanks to Tim Adye it can be used also through
\texttt{RooUnfold}.

\end{document}